# COUPLING CURRENT CONTROL IN STABRITE-COATED NbTi RUTHERFORD CABLES BY VARYING THE WIDTH OF A STAINLESS STEEL CORE

M.D. Sumption[1], E.W. Collings[1], A. Nijhuis[2], and R.M. Scanlan[3]

[1]LASM, MSE, The Ohio State University
Columbus, OH 43210, USA
[2]The University of Twente,
Enschede, 7500 AE, The Netherlands
[3]Superconducting Magnet Group, LBNL
Berkeley, CA 94720, USA

## ABSTRACT

AC loss measurements were made on a series of Rutherford cables wound from stabrite-coated NbTi/Cu multifilamentary strand, from the results of which interstrand contact resistances were calculated. The 28-strand, 15 mm wide, LHC-inner type cables contained stainless steel cores of widths varying from 0 (no core) to 12.7 mm. Measurements were made on 4-layer cable packs. Two main preparation (curing) cycles were used: (1) heat treatment (HT) at 170°C under 80 MPa followed by pressure release and re-application before measurement; (2) HT to 200°C and maintenance of the pressure until after measurement. One additional protocol-(1)-treated pack was prepared from cables that had received a diffusion HT of 200°C/8h in air following a CERN-recommended prescription for achieving a satisfactory contact resistance (ICR). Calorimetric loss measurements were made on all samples using a sinusoidal field with an amplitude of 400 mT. It is concluded that by adjusting the width of a stainless steel core the effective ICR, $R_{\perp,eff}$, of a stabrite-coated LHC-inner-type Rutherford cable can be varied over a wide range, and in particular that a published target of 15 µΩ could be attained with a core width of 8.5 mm – a little more than one-half the width of the cable, leaving the rest of the cable free for current sharing. It is also concluded that in terms of $R_{\perp,eff}$ the OSU-administered CERN diffusion HT was equivalent to inserting a 9.5-mm-wide core.

## INTRODUCTION

In Rutherford cables interstrand contact resistance, ICR, is an important parameter that controls cable stability as well as the magnitude of interstrand coupling currents with their attendant parasitic magnetic fields. In most NbTi/Cu-wound Rutherford cables ICR is determined by the resistivity of the native oxide layer on the strand surface [1-9] which may be that of the Cu itself or that of, for example, a Cr or stabrite coating. In some cases the surfaces are specially oxidized or given other non-metallic coatings which lead to different characteristic ICRs. Stabrite coated cables are a special case; initially used in the virgin condition, they are now given a 200°C heat treatment during which some Cu diffuses through the stabrite to form an oxidized layer on its surface (in which case the ICR may also partly depend on the thickness of the stabrite coating[8]).

In addition to strand surface condition, the ICR of a given cable also depends strongly on the cable preparation conditions – "curing" time, temperature, and pressure[1-11]. Beyond this, pressure during measurement as well as (in some cases) pressure cycling are important modifiers of the ICR [1,3,4,6-13].

The ICR must be set sufficiently high to suppress interstrand coupling currents, but sufficiently low to ensure adequate current sharing between strands. For the LHC inner winding optimal ICR is estimated to be 15 to 20 $\mu\Omega$ [6,7]. As indicated above, meticulous control of strand surface and cable preparation conditions should lead to an acceptable ICR. With such an approach reproducibility is an obvious difficulty. On the other hand a more direct approach, in that it involves fewer variables, is to moderate ICR through the insertion of a core. Although most any material will serve the function of a core, stainless steel AISI 316 has turned out to be generally satisfactory.

Following their introduction some twenty years ago[14] resistive cable cores have been the subject of detailed study[4,5,15-19]. The ICRs of cored and uncored cables have been compared, particular attention being given to curing conditions, sample preparation protocols, and measurement conditions. In other recent experiments the influences on ICR of core thickness variations and cable compaction have also been studied[20]. As usual, the core is responsible for a drastic reduction of coupling (current) loss, while at the same time increases in core thickness or the application of compaction (uniaxial stress) serve to increase side-by-side interstrand contact (lowering the $R_{//}$ component of ICR), and hence slightly *elevate* the loss. The preservation or lowering of $R_{//}$ helps to maintain cable stability at low-to-moderate values of $I/I_c$. On the other hand, current sharing at $I/I_c$ close to 90% relies on a low value of the crossover ICR, $R_\wedge$. A full-width core, while it completely suppresses crossover contact and drastically lowers coupling magnetization and loss, may at the same time be harmful to cable stability. For this reason it is important to know just how far the core width can be reduced while still preventing the effective ICR from decreasing below the above-mentioned 15-20 $\mu\Omega$.

**EXPERIMENTAL**

The cables prepared for these experiments were of the LHC-dipole, inner-winding type, with 28 strands, a keystone angle of about 1.2-1.25°, and a transposition pitch of 115 mm. These 15-mm wide cables were wound at the Lawrence Berkeley National Laboratory with 1.07 mm diam. stabrite-coated NbTi/Cu mulitfilamentary strand (Alsthom) and provided with 0.001 in. (25 µm) thick AISI-316 stainless steel cores of widths 0 (no core), 1/8, 1/4, 3/8, and 1/2 in. (3.2, 6.4, 9.5, and 12.7 mm). These cables were designated A, B, C, D, and E, respectively.

Prepared for measurement were 4-high packs of cable 40 cm in length. Two sets of such packs were prepared: (1) A **PR** ("pressure-release") set in which kapton- plus B-stage-epoxy-wrapped cables were torqued down in a 36-bolt fixture to a pressure of 80 MPa and then heat treated (HT) in a simple box furnace set at 170°C. A mounted cable pack was left in the furnace for five hours (after which it had attained a temperature of 168°C), removed and air cooled. For measurement the "cured" pack was returned to the fixture and repressurized to 80 MPa (room temperature). These packs were designated A170, B170, C170, D170, and E170, respectively. (2) A **CP** ("constant pressure") set in which kapton-wrapped cables (no epoxy needed this time) were torqued down in the 36-bolt fixture to 80 MPa and given an HT consisting of a 1 h ramp to 200°C followed by a 2 h hold at that temperature. The mounted cables were transferred to the cryostat for measurement without release of the pressure. These packs were designated A200, B200, C200, D200, and E200, respectively.

The processing schedule of the **PR** cable packs was intended to mimic that of a typical accelerator dipole winding. The relatively extreme **CP** procedure was intended to emphasize the properties of a variable width core in the environment of a fully sintered cable.

Two more cable packs were prepared for measurement: (1) Sample CN170 prepared from uncored (A-type) cable that had been heat treated at 8h/200°C in air while being completely boxed-in by "sacrificial" lengths of the same type of cable. This HT was intended to simulate the CERN-recommended annealing of entire spools of stabrite-coated cable, the purpose of which was to engineer a predetermined ICR by allowing some Cu to diffuse out to the strand's stabrite surface and to oxidize there. A cable pack was then prepared following Protocol-PR above. (2) A pack of fully-cored E-type cables were mounted in the holder under minimal pressure and not heat treated; measurement of this reference sample, designated REF, would provide an intrastrand-only eddy current loss baseline. After pressurization and HT, this pack was later measured as E200.

Calorimetric measurements of AC loss were performed at the University of Twente using equipment previously described[13]. The measurements were performed at 4.2 K in an AC field of amplitude $B_m$ = 400 mT at frequencies between 10 and 90 mHz.

**ANALYSIS**

Based on equations listed by Sytnikov et al. [21,22] for coupling power loss per unit length of a Rutherford cable exposed to a ramping magnetic field $B$ (of amplitude $B_m$) and applied either perpendicular ($\perp$, "face on", FO) or parallel ($\parallel$, "edge on", EO) to the broad face of the cable is the following expression for energy loss per cycle per m$^3$ of outside cable dimensions (width $w$ and thickness $t$)

$$Q_\perp = \frac{4}{3}\left(\frac{w}{t}\right)L_p B_m \left(\frac{dB}{dt}\right)\left[\frac{N^2}{20R_\perp} + \frac{1}{NR_\parallel}\right] \quad (1)$$

and

$$Q_\parallel = \left(\frac{t}{w}\right)L_p B_m \left(\frac{dB}{dt}\right)\left[\frac{1}{NR_\parallel}\right] \quad (2)$$

where $N$ is the number of strands and $L_p$ is 1/2 the transposition pitch of the winding (and $L_s$ is the corresponding length of strand). In Eq. (1) $R_\perp$ is the resistance of each interstrand crossover contact and $R_\parallel$ is the side-by-side resistance (per length $L_s$) between adjacent pairs of strand. For general cable-to-cable comparison it is useful to lump the effects of all contact resistances into an "effective interstrand contact resistance", $R_{\perp,eff}$ defined by

$$Q_\perp = \frac{4}{3}\left(\frac{w}{t}\right)L_p B_m \left(\frac{dB}{dt}\right)\frac{N^2}{20R_{\perp,eff}} \quad (3)$$

In order to apply these equations to our samples we have normalized the measured loss to total cable volume (consistent with the normalization of the above expressions).

Finally, we note that the loss equations listed above are valid only for linearly ramping fields (e.g., triangular and trapezoidal waves). In the present case of a sinusoidal field it is useful[23] to use the rms average of $dB/dt$, namely , $<|dB/dt|>_{rms.} = (\pi^2/8)^{1/2} 4fB_m$, and it is necessary to attach another factor of $(\pi^2/8)^{1/2}$ as a prefactor to the above loss equations, after which Eq. (3) becomes

$$Q_\perp = \frac{2}{3}p^2\left(\frac{w}{t}\right)L_p B_m^2 \frac{N^2}{20 R_{\perp,eff}} f \qquad (4)$$

Equation (4) shows that $R_{\wedge,eff}$ may be obtained from the (reciprocal) slope of $Q_\wedge$. In practice the $f$-dependent component of the total cable loss, $Q_{tot}$, is the sum of the coupling and strand-eddy-current ($Q_e$) components. A separate experiment performed on an uncompacted stack of fully cored cables yields the eddy current component, thereby enabling the separation of $Q_\wedge$ from the measured $Q_{tot} = Q_\wedge + Q_e$.

**RESULTS**

The direct results of the measurements of the **PR** and **CP** series cable packs in terms of $Q_{tot}$ vs frequency are presented in Figs. 1(a) and 1(b), respectively. After adjustment for the intrastrand eddy current loss, the FO coupling losses within the initial linear segments of $Q_\wedge(f)$ to which the standard loss equations apply are presented in two formats: (1) Fig. 2 which shows in normalized terms the rate at which $Q_\wedge$ drops as the width of the core increases; (2) Fig. 3, based on the previous figure, which depicts the core-width dependence of the Eq.(4)-calculated values of $R_{\wedge,eff}$.

Figs. 1(a) and 1(b) both show the losses decreasing rapidly as the core width increases and crossover contact becomes more and more suppressed. As expected, in the fully sintered

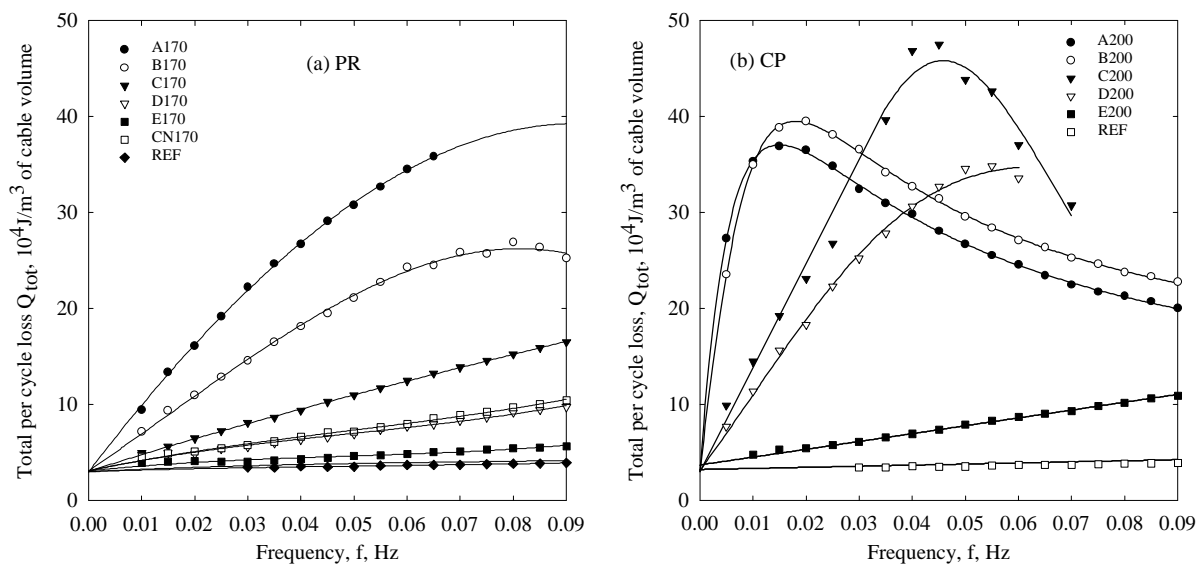

Figure 1. Total per cycle loss for (a) the PR-series and (b) the CP-series samples.

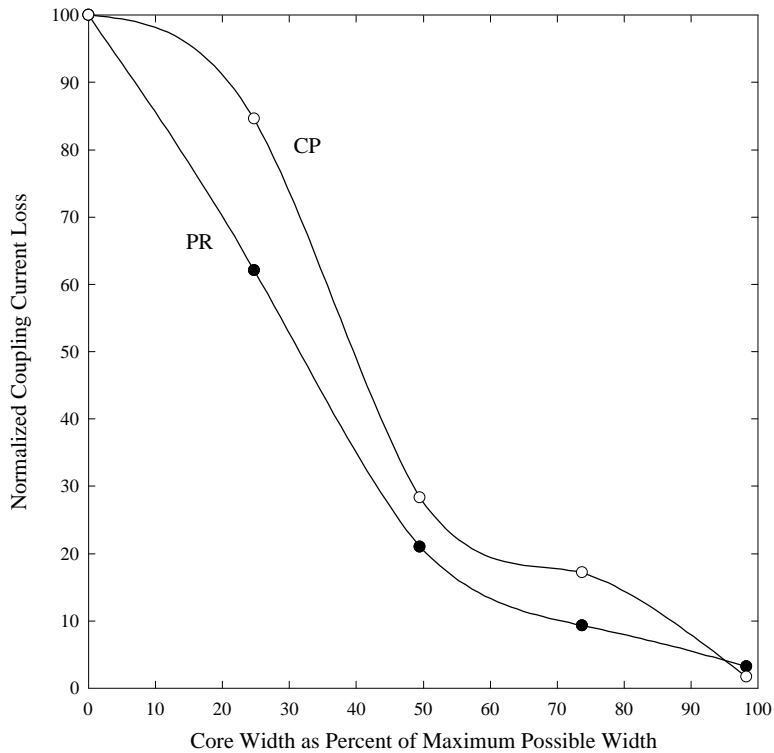

Figure 2. FO Coupling loss, $Q_\wedge$, normalized to its no-core value at the same frequency for the initial linear segment of $Q_\wedge(f)$.

**CP**200 series the interstrand contact is much stronger than it is in the more practical **PR**170 case, so much so that the critical frequency ($f_c$, at the maximum of $Q_{tot}$) falls within the frequency range of the measurements. For the engineer interested in AC loss as such, Fig. 2 indicates that as the core width, $w_{core}$, increases up to $w_{mav}/2$, where $w_{mav}$ is the maximum available core width, the normalized loss drops by about 70% of its initial value. On the other hand if $R_{\wedge,eff}$ is the principal focus, Fig. 3 (essentially the reciprocal of the earlier figure) shows an initially small $R_{\wedge,eff}$ to increase rapidly with $w_{core}$ beyond some critical values, viz. $w_{mav}/2$ in the case of the **PR**170 cables and 75% of $w_{mav}$ in the case of the **CP**200 series.

## CONCLUDING DISCUSSION

### The CP Results

As mentioned, the **CP** procedure was intended to explore the properties of a variable width core in the environment of a fully sintered cable. In such a cable, the ICR is so low that the core needs to cover more than 75% of the available area before an acceptable $R_{\wedge,eff}$ (e.g. 5-10 $\mu\Omega$) can develop. For most of the **CP** cables the $f_c$s are sufficiently low that $Q_{tot,max}$ appears within the frequency range of the measurements, Fig. 1(b). As coupling currents become suppressed $Q_{tot,max}$ is expected to decrease with increasing $w_{core}$. Although this seems to be the case for the **PR** cables, with the **CP** cables the insertion of cores initially increases $Q_{tot,max}$. It is postulated that for the **CP** series a strong variation of ICR across the width of the cable stimulates an additional "boundary induced coupling current" or BICC[24]. In the unsintered **PR** cables such a variation would be weaker and no anomalous $Q_{max}(f)$ trend is expected, cf. Fig. 1 (a).

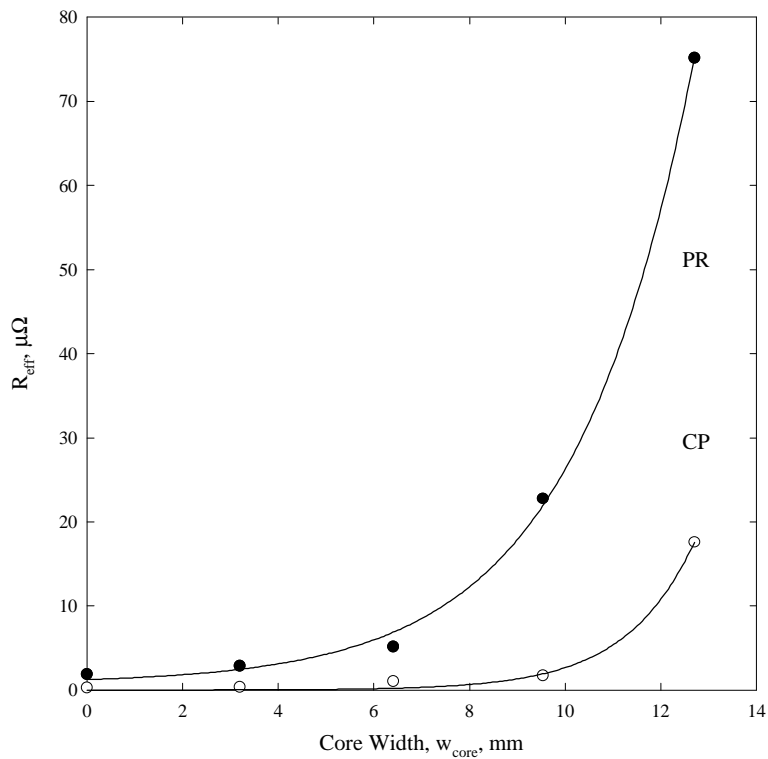

Figure 3. $R_{\wedge,eff}$ vs core width.

### The PR Results

The **PR** protocol is of greater interest to the magnet designer since in practice a pressure-cured winding is transferred after pressure release to the magnet body in preparation for repressurization during collaring and in service. Figure 3 indicates that under **PR** conditions the desired $R_{\wedge,eff}$ = 15-20 µΩ can be achieved by increasing the $w_{core}$ up to about 65-70% $w_{mav}$. Clearly a reduced width core is preferable to a full width one in that it permits some 30-35% of the cable width to be available for clean low-ICR crossover-type current sharing. CN170 is a (core-free) pack that had been prepared from previously diffusion-HT (8h/200°C) cables. For it, Fig. 1(a) indicates an $R_{\wedge,eff}$ a little less that that of D170 (23 µΩ).

### ACKNOWLEDGEMENTS

The cables were wound at the Lawrence Berkeley National Laboratory by H. Higley, and the research was supported by the U.S. Department of Energy, Division of High Energy Physics, under Grants No. DE-FG02-95ER40900 (OSU) and DE-AC03-76SF00098 (LBNL).